# Continuously Tunable Berry Phase and Valley-Polarized Energy Spectra in Bilayer Graphene Quantum Dots

Ya-Ning Ren[1,§], Qiang Cheng[2,3,§], Qing-Feng Sun[3,4,5,*], and Lin He[1,*]

[1]Center for Advanced Quantum Studies, Department of Physics, Beijing Normal University, Beijing, 100875, People's Republic of China

[2]School of Science, Qingdao University of Technology, Qingdao, Shandong 266520, China

[3]International Center for Quantum Materials, School of Physics, Peking University, Beijing, 100871, China

[4]Collaborative Innovation Center of Quantum Matter, Beijing 100871, China

[5]Beijing Academy of Quantum Information Sciences, West Bld. #3, No. 10 Xibeiwang East Road, Haidian District, Beijing 100193, China

[§]These authors contributed equally to this work.
*Correspondence and requests for materials should be addressed toQing-Feng Sun (e-mail: sunqf@pku.edu.cn) and Lin He(e-mail: helin@bnu.edu.cn).

**Berry phase plays an important role in determining many physical properties of quantum systems. However, a Berry phase altering energy spectrum of a quantum system is comparatively rare. Here, we report an unusual tunable valley polarized energy spectra induced by continuously tunable Berry phase in Bernal-stacked bilayer graphene quantum dots. In our experiment, the Berry phase of electron orbital states is continuously tuned from about $\pi$ to $2\pi$ by perpendicular magnetic fields. When the Berry phase equals $\pi$ or $2\pi$, the electron states in the two inequivalent valleys are energetically degenerate. By altering the Berry phase to noninteger multiples of $\pi$, large and continuously tunable valley polarized energy spectra are detected in our experiment. The observed Berry phase-induced valley splitting, on the order of 10 meV at a magnetic field of 1 T, is about 100 times larger than Zeeman splitting for spin, shedding light on graphene-based valleytronics.**

In two-dimensional honeycomb lattice systems with broken spatial inversion symmetry, the Berry curvature in two inequivalent valleys has opposite signs, which enables the control of the valley degree of freedom [1-10]. Among various material candidates for valleytronics, gapped Bernal-stacked bilayer graphene (BLG) is one of the most studied systems, showing great promise in terms of tunability of the valley current and valley splitting [6-19]. The opposite Berry curvature and the associated magnetization of electron states in the two valleys of the gapped BLG lead to a large linear magnetic field splitting of the valley degeneracy. Recently, the Berry curvature-induced valley splitting has been demonstrated by single-carrier measurements (in the Coulomb blockade regime) in BLG-based quantum dot (QD) devices [11-15,18]. Besides the Berry curvature effects, it was also predicted to realize valley polarized bound states in BLG QDs when the Berry phase accumulated in the two valleys is tuned to noninteger multiples of $\pi$ [19]. Yet, in most cases studied to date, the Berry phase in graphene systems equals to integer multiples of $\pi$ [20-25] and altering the Berry phase to noninteger multiples of $\pi$ is experimentally challenge. Therefore, the direct observation of the Berry phase-induced valley-polarized energy spectrum in the BLG QDs has remained elusive.

In this Letter, the Berry phase of electron orbital states in the BLG QDs is continuously tuned from about $\pi$ to $2\pi$ by perpendicular magnetic fields and unusual Berry phase-induced valley-polarized energy spectra are observed. In our experiment, a scanning tunneling microscope (STM) tip is used to approaching the BLG to introduce a movable confining potential *i.e.*, a QD, in the BLG below the tip [4,5,13,25,26]. Clear resonances due to the formation of bound states are observed in the BLG QD. By applying a perpendicular magnetic field, the Berry phase of the bound states is continuously tuned from about $\pi$ to $2\pi$. When the Berry phase is not equal to $\pi$ or $2\pi$, valley-polarized energy spectra with tunable valley splitting are observed.

The Berry phase is the flux of the Berry curvature integrated over the area circled by the closed path in the momentum space. In graphene QD, a magnetic field enables fine control of the trajectories and hence the Berry phase for individual confined states [18,26], as summarized in Fig. 1 (see Supplemental Material for details [28]). Therefore, graphene QD offers an ideal platform to study the effect that the energy spectrum is tuned by the Berry phase. We should point out that the results obtained in monolayer graphene QD and the BLG QD are quite different. For monolayer graphene, the Berry phase accumulated in the two valleys jumps from 0 to $\pm\pi$ at a critical value of the magnetic field because that the Berry curvature is only nonzero at the Dirac point: the Berry phase of the quasibound states is zero when their corresponding momentum-space loop does not enclose the origin, and the Berry phase becomes π when the momentum-space loop encircles the Dirac point (Figs. 1(b) and 1(c)). The π shift of the Berry phase will suddenly lift the degeneracy of the quasibound states with opposite angular momenta $\pm m$ (the quasibound states in the two valleys are still energetically degenerate), as demonstrated recently in Ref. [25,29,30]. For the BLG, the Berry curvature is ring-shaped in the momentum space (Fig. 1(e)). Then, the area circled by the closed path and, simultaneously, the Berry phase can be continuously tuned by the magnetic field (Fig. 1(f)). When the Berry phase becomes noninteger multiples of π, the Berry phases of the two valleys are inequivalent and the energy of the bound states for the two valley becomes different according to the Einstein-Brillouin-Keller (EBK) quantization rule [19]. Therefore, valley degeneracy of the bound states is lifted and an unusual valley-polarized energy spectrum can be obtained (see Supplemental Material for details [28]).

Our experiments were carried out on decoupled Bernal-stacked BLG on graphite [9,23,31,32] by using a high-magnetic-field STM at $T$ = 4.2 K (see Supplemental Material for details [28]). The decoupled Bernal-stacked BLG is identified by both the STM image and the scanning tunneling spectroscopy (STS) spectra (Fig. 2). The atomic-resolution STM image exhibits a triangular lattice (Fig. 2(a)), arising from the A/B atoms' asymmetry in the Bernal-stacked BLG. The high-magnetic-field STS spectra show well-defined Landau quantization of massive Dirac fermions (Figs. 2(b) and 2(c)), which demonstrates explicitly that the studied system is decoupled Bernal-stacked BLG [9,23,31,32] (see Supplemental Material Fig. S1 for details of analysis [28]). In zero magnetic field, the tunneling spectrum exhibits a pronounced peak at the edge of the conduction band (Figs. 2(b), 2(f) and Fig. S2 [28]), indicating emergence of a flat band in the Bernal-stacked BLG. Such a feature was also observed in STS spectra of the decoupled Bernal-stacked BLG on graphite in the literature [31,32], however, it has been ignored in discussions so far. Recently, angle-resolved photoemission measurements for Bernal-stacked BLG on SiC demonstrate the formation of the flat band in the BLG when there is sublattice asymmetry in the bottom layer due to the interaction of the substrate [33]. In our experiment, the graphite substrate introduces interlayer asymmetry (hence the gap in the BLG, but the potential difference between top and bottom layers are much less than the nearest interlayer hopping energy $t_\perp$), leading to the flat band. According to our experimental result and calculation, the band structure is extremely flat, showing a 0.084-meV dispersion, around the conduction band edge (Figs. 2(d)-(f)). Therefore, the full-width at half-maximum (FWHM) of the flat band in the BLG measured in our experiment is comparable to that in magic-angle twisted bilayer graphene (MATGB) (also measured by using STM) [34-39].

To introduce the BLG QDs in our study, we used a STM tip as a top gate to generate band bending of the BLG below the tip [4,5,13,25,26]. In the experiment, the STM tip is approaching the BLG by increasing the tunneling current with a fixed voltage bias, as shown in Fig. 3(a). For short tip-sample distance, the signal of the flat band is further enhanced in the tunneling spectra. Besides that, the work function difference between the STM tip and the BLG leads to an effective electric field acting on the BLG and results in the confining potential on the massive Dirac Fermions. Then, besides the flat band, several almost equally-spaced resonances, which are attributed to the confined bound states in the BLG QD, are observed in the STS spectrum (Figs. 3(b) and 3(c)). In our experiment, the process shown in Fig. 3(a) can be reproducible obtained in any selected position of the decoupled BLG and the energy spacing of the resonances in the tip-induced BLG QD is almost the same and independent of the measured positions (see Fig. S3 for more data [28]).

To explore the Berry phase-induced valley-polarized energy spectra in the BLG QD, we carried out STS measurements in magnetic fields with a small interval of the magnetic field $\Delta B = 0.05$ T, as shown in Fig. 4(a) (left panel). With increasing magnetic fields from 0 T, a notable splitting of the bound states can be observed. At 1 T, the splitting is about 10 meV, which is about 100 times larger than that of Zeeman splitting for spin. When the magnetic field increases to about 3 T, two adjacent split states merge into a new state. Such a feature reminds us the Berry phase-induced valley polarized energy spectra in the BLG QDs [19]. The continuously changing of the Berry phase for the BLG leads to unusual behaviors of the valley-related energy spectrum. According to the semiclassical EBK quantization rule [19,27], one has

$$\oint_{C_r} \Pi_r \, dr = 2\pi \left(n + \frac{1}{2}\right) + \gamma, \qquad (1)$$

for the valley $K$ and

$$\oint_{C_r} \Pi_r \, dr = 2\pi \left(n + \frac{1}{2}\right) - \gamma, \qquad (2)$$

for the valley $K'$ with the integer $n$. Here the Berry phases for the valleys $K$ and $K'$ are opposite. The left sides of Eqs. (1) and (2) are dimensionless. Since $\Pi_r$ (the radial momentum) and the Berry phase $\gamma$ are functions of energy $E$, the Eqs. (1) and (2) determine the behaviors of the valley-related bound states in the QDs (see Fig. S4 and Supplemental Material for details of calculation [28]). When the Berry phase $\gamma = 0, \pi$ or $2\pi$, the bound states for the valleys $K$ and $K'$ are degenerate. In other situations, the bound states for the valleys $K$ and $K'$ are split. This is essentially different from the monolayer graphene, in which the Berry phase only is 0 or $\pi$ and the bound states in two valleys are always degenerate. As a result, when the Berry phase $\gamma$ of the BLG changes continuously from $\pi$ to $2\pi$ by increasing the magnetic field from zero, the energy spectrum of the bound states will experience the unusual degenerate-splitting-degenerate process of the valley degree of freedom, as directly observed from the STS spectra of the BLG QD (Fig. 4(a) (left panel)).

To further understand the Berry phase-induced valley-polarized energy spectra in the BLG QD, we calculated the local density of states (LDOS) for the BLG QD fully based on the quantum mechanics. For simplicity, we model the BLG QD by the Hamiltonian $\widetilde{H}_\xi = H_\xi + U(r)$, where $U(r) = \kappa r^2$ is the parabolic potential with the strength $\kappa$. $H_\xi$ is the $4 \times 4$ Hamiltonian for the ungated BLG

$$H_\xi = v\tau_0(\xi\Pi_x\sigma_x + \Pi_y\sigma_y) + \frac{t_\perp(\tau_x\sigma_x+\tau_y\sigma_y)}{2} + \frac{\Delta_1(\tau_0+\tau_z)\sigma_0}{2} + \frac{\Delta_2(\tau_0-\tau_z)\sigma_0}{2}, \quad (3)$$

in the layer⊗sublattice space, with the interlayer hopping energy $t_\perp$, the potentials $\Delta_1$ and $\Delta_2$ of the top and bottom layers, the valley index $\xi = \pm 1$, the Fermi velocity $v$, the Pauli matrices $\boldsymbol{\sigma} = (\sigma_x, \sigma_y)$ and $\boldsymbol{\tau} = (\tau_x, \tau_y)$ in the sublattice space and the layer space, the momentum $\boldsymbol{\Pi} = (\Pi_x, \Pi_y) = (-i\hbar\partial_x - eA_x, -i\hbar\partial_y - eA_y)$ with the vector potential $\boldsymbol{A} = (A_x, A_y) = B(-y, x, 0)/2$. In the BLG QD with the rotational symmetry, the LDOS at $r = r_0$ can be expressed as $D(E) = \sum_M D_M(E)$ with $D_M(E)$ being the LDOS contributed from the angular momentum $M$-state [19,27] (see Supplemental Material for details [28]). In order to explicitly show the effects of the continuous change of the Berry phase, we calculated the LDOS at the center of the BLG QD with $r_0 = 0$ for $M = 0$ and $\pm 1$ in Fig. S5 [28]. Figure 4(b) shows the result for $M = 0$ as an example. Both the LDOS of the top layer and the bottom layer are included. The bound states of the two valleys are degenerate for the negative magnetic field with large enough absolute value due to the Berry phase $\gamma = 0$. When the field is increased, the bound states of the two valleys start to split due to the finite value obtained by the Berry phase. When the Berry phase reaches to the value $\pi$, the bound states become degenerate again. The bound states of valley $K$ and that of $K'$ cross each other at $\gamma = \pi$. As the field is continuously raised, the crossing lines of the valley-related states split again because of $\gamma > \pi$. When the Berry phase achieves the value of $2\pi$ for lager enough positive magnetic field, valley degeneracy happens again. Finally, the levels of the valleys $K$ and $K'$ are recombined into degenerate Landau levels when the magnetic length is smaller than the effective radium of the QD. The switching processes of degeneracy-splitting for levels embodied in the LDOS are consistent with the behaviors of the continuously changed Berry phase. For other values of $M$, these characteristics of the LDOS still exist.

In our experiment, the on-center STM measurement mainly reflects the LDOS of the top layer of the BLG. Therefore, we present the numerical result of top-layer

LDOS at the center of the BLG QD, as shown in Fig. 4(a) (right panel). The contributions from $-10 \leq M \leq 10$ are considered. Although the switching processes of degeneracy-splitting are no longer concise compared with those in Fig. 4(b), the continually tunable Berry phase is still responsible for the peculiar features of the LDOS. The numerical calculation reproduces quite well the main features of the experimental result, which provides strong proof of the Berry phase-induced valley-polarized energy spectra in the BLG QD. In our experiment, the flat band is almost independent of the magnetic fields and there is a large energy separation between the flat band and the lowest bound state (Fig. 4(a) (left panel)). There are two possible scenarios for such a feature. One is that the STM tip cannot generate sufficient band bending to form the bound states of the flat band because of its large LDOS. Therefore, there is a large energy separation between the flat band and the bound states confined from other electron bands. The other is that such a large energy separation may arise from electron-electron interactions, as observed in the flat bands of the MATGB [35-38,40]. The above two effects are not taken into account in our theoretical calculations.

In summary, large and tunable valley polarized energy spectra are realized in the BLG QDs by continuously tuning the Berry phase with magnetic fields. Our results demonstrate the close relationship between the valley polarization and the noninteger multiples of $\pi$ of the Berry phase, which reveals the Berry phase's essential role in valleytronics.

**Acknowledgments**

This work was supported by the National Natural Science Foundation of China (Grant Nos. 11974050, 11674029, 11921005) and National Key R and D Program of China (Grant No. 2017YFA0303301). L.H. also acknowledges support from the National Program for Support of Top-notch Young Professionals, support from "the Fundamental Research Funds for the Central Universities", and support from "Chang Jiang Scholars Program".

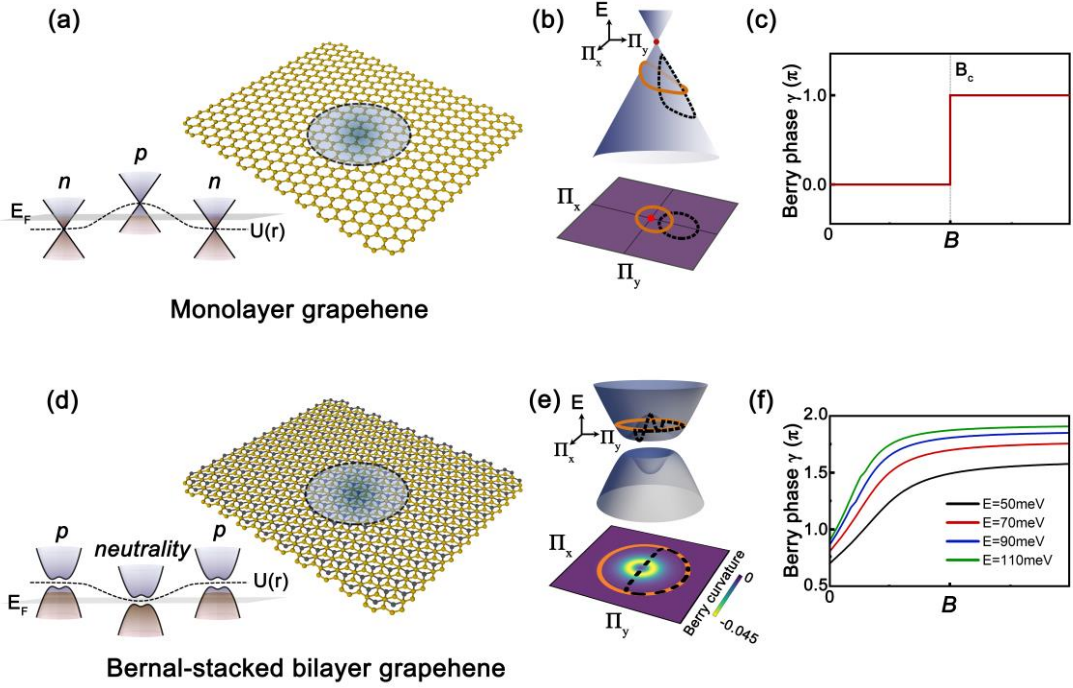

**FIG. 1.** **(a)-(c)** Jump of the Berry phase in monolayer graphene QDs. **(e)-(h)** Continuously tunable Berry phase in Bernal-stacked BLG QDs. **(a)** and **(d)** Sketches of monolayer graphene QD and BLG QD, respectively. **(b)** and **(e)** Schematic charge trajectories in momentum space of monolayer graphene QD and BLG QD, respectively. The orange solid and black dashed lines in panel (b) represent trajectories in the magnetic fields above and below the $B_C$, respectively. The orange solid and black dashed lines in panel (e) represent trajectories in the magnetic fields of $10T$ and $0T$, respectively. **(c)** and **(f)** Berry phase as a function of the magnetic fields $B$ in monolayer graphene QD and BLG QD, respectively.

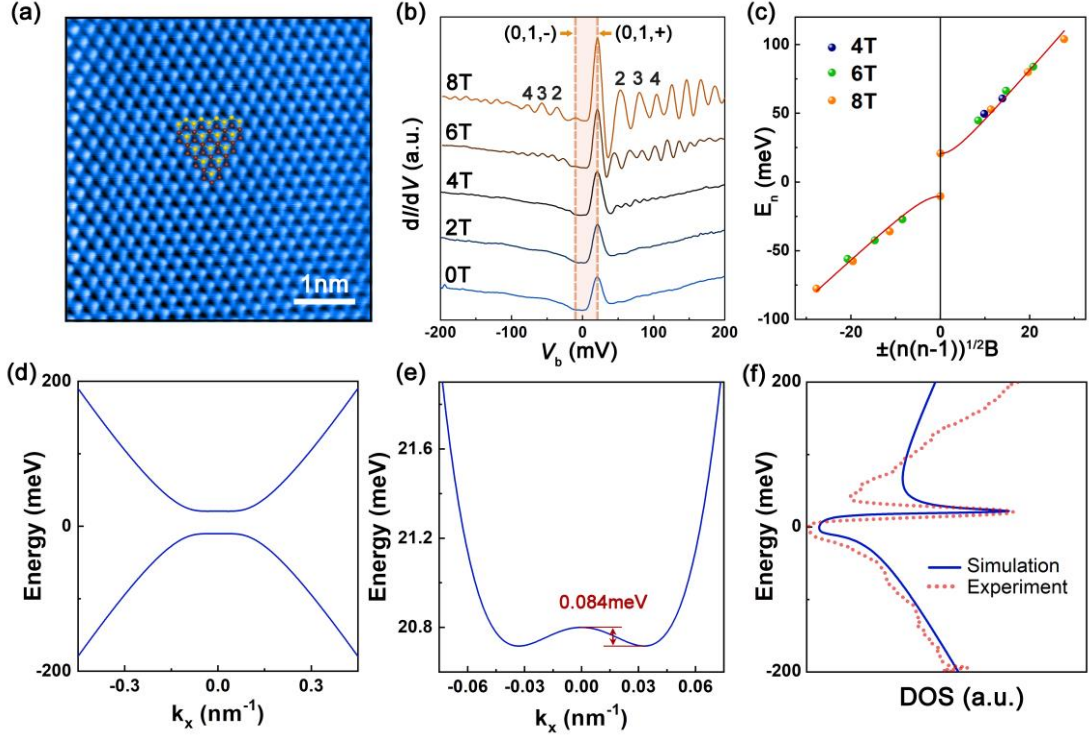

**FIG. 2. (a)** A 5 ×5 nm² atomic resolved STM image ($V_{\text{sample}}$ = 800 mV, $I$ = 200 pA) of the Bernal-stacked BLG. The triangular graphene lattices of the Bernal-stacked BLG are overlaid onto the image. **(b)** Landau level spectra of the BLG for various magnetic fields. Curves are shifted vertically for clarity. The Landau level peak indices are marked and the gap is labelled by shadows. **(c)** The Landau level energies for different magnetic fields obtained from panel (b) against $\pm(n(n-1))^{1/2}B$. **(d)** Calculated low-energy dispersions of the Bernal-stacked BLG. **(e)** A zoom-in image of the calculated flat band dispersion in panel (d). **(f)** Experimental and calculated DOS of the top layer as a function of energy.

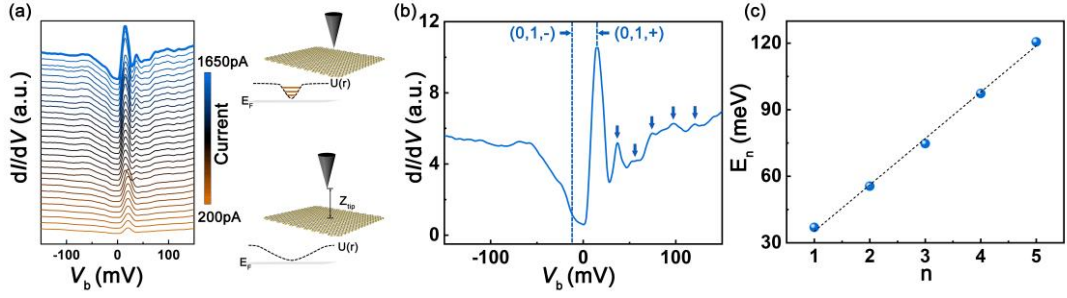

**FIG. 3. (a)** The dI/dV spectra measured at different tip-sample distances $Z_{tip}$. The tip height decreases by increasing the tunneling current with a fixed voltage bias. Curves are shifted vertically for clarity. When the tip is close enough to the BLG, the electric field of the STM tip will induce band bending to confine the charge carriers, as schematically shown in the right panel. **(b)** The *dI/dV* spectrum acquired at *I*=1650 pA in panel (a). The bound states generated in the tip-induced BLG QD are marked by arrows. **(c)** The bound states generated in the tip-induced BLG QD show equally energy spacing distribution.

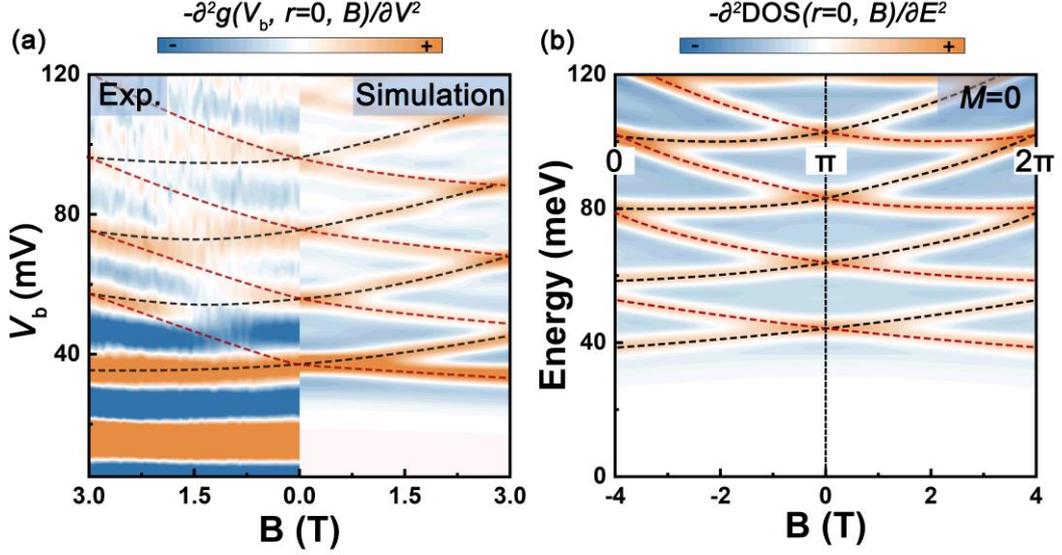

**FIG. 4. (a)** Experimental (left panel) and calculated (right panel) differential conductance maps versus magnetic fields $B$ in the center of the BLG QD. Here calculated differential conductance is proportional to the differential LDOS $-\partial^2 D(E)/\partial E^2$ of the top layer. The black (red) dashed lines guide the trend of bound states for the valley $K(K')$. **(b)** The LDOS in the center of the BLG QD for $M = 0$, showing the effects of the continuous changes of the Berry phase on the valley-related bound states. The black lines guide the trend of levels for the valley $K$. The red lines guide the trend of levels for the valley $K'$.